\documentclass[twocolumn,showpacs,preprintnumbers,amsmath,amssymb]{revtex4}

\usepackage{graphicx}
\usepackage{dcolumn}
\usepackage{bm}
\usepackage{subfigure}

\begin{document}


\title{Real space Green's function approach to angle resolved resonant photoemission: \\
 spin polarization and circular dichroism in itinerant magnets}

\author{Fabiana Da Pieve$^{1}$} \email{fabiana.dapieve@gmail.com}
\author{Peter Kr\"uger$^{2}$}

\affiliation{
$^1$ ALGC, Vrije Universiteit Brussel, Pleinlaan 2, Brussels 1050, Belgium\\
$^2$ Graduate School of Advanced Integration Science, Chiba University, 1-33 Yayoi-cho, Inage, Chiba 263-8522, Japan\\
}

\date{\today}

\begin{abstract}

A first principles approach, based on the real space multiple scattering Green's function method, is presented for spin- and angle-resolved resonant photoemission from magnetic surfaces. It is applied to the Fe(010) valence band
photoemission excited with circularly polarized X-rays around the Fe $L_3$
absorption edge. When the photon energy is swept through the Fe $2p-3d$
resonance, the valence band spectra are strongly modified in terms of
absolute and relative peak intensities, degree of spin-polarization
and light polarization dependence.
New peaks in the spin-polarized spectra are identified as
spin-flip transitions induced by exchange decay of spin-mixed
core-holes. By comparison with single atom and band
structure data, it is shown that both intra-atomic and multiple
scattering effects strongly influence the spectra. We show how the different features linked to states of different orbital symmetry in the $d$ band are differently enhanced by the resonant effect. The appearance and origin of circular dichroism and spin polarization
are analyzed for different geometries of light incidence and electron
emission direction, providing guidelines for future experiments.

\end{abstract}	

\begin{description}
\item[PACS numbers]
\end{description}

\pacs{78.20.Bh, 78.20.Ls,78.70.-g,79.60.-i}
\maketitle


In the last decade, magnetic circular dichroism (MCD) and spin polarization studies in resonant
inelastic X-ray scattering (RIXS) and resonant photoemission (RPES) have acquired great importance in the study of magnetic and correlated materials. Such spectroscopies probe respectively the radiative and non radiative autoionization decay of a core hole, and the signal can be strongly enhanced with respect to the non resonant mode. The element and orbital selectivity of core level resonant spectroscopies allows to access higher order multipoles which are left unexplored by MCD in X-ray absorption (XAS) \cite{quadr,supersumrules,mcdnormal2,prb69,multi1,multi2}, to distinguish and enhance specific
electronic excitations and satellites \cite{prb46,magnuson}, collective magnetic excitations \cite{haverkort}, ultrafast and charge transfer dynamics \cite{ultra,charge1,charge2} and to detect quadrupolar transitions towards localized empty states \cite{loc1,loc2}. In particular, RPES has recently been applied to several correlated materials \cite{sawatzky,pauliu,bapna,oht,koit} and full two dimensional angular scans of resonantly emitted electrons in moderately correlated materials have also been carried out \cite{peter,greber}. These works, together with earlier pioneering studies \cite{tjeng,sinkovic} on local magnetic properties in macroscopically non magnetic systems, demonstrate the importance of RPES and the need for an advancement in the theoretical description of this spectroscopy, which is the main aim of this work. 

RPES is in principle an autoionization channel of the more general
process called resonant Auger decay. Depending on whether the core-excited electron participates or not in the Auger decay, the process is termed either participator or spectator channel. In the participator channel the one hole
final state is degenerate with the one in direct valence band
photoemission (PES, or ARPES if angle resolved). Thus the
two processes generally interfere, giving rise to a typical Fano profile
\cite{fano} and the emission is often strongly enhanced. This autoionization
channel shows linear dispersion of the spectral features with
photon energy (Raman regime), as direct PES, and it is the one that, strictly speaking, constitutes the RPES. The spectator
channel, on the contrary, leads to an Auger-like final state with two holes, and the spectral
lines exhibit a normal Auger behaviour.
However, often the enhancement of the direct valence band PES is given by the combination
of the two different photoemission-like and Auger-like channels \cite{mart,alm,dru,charge2,sawatzky} and, in order to
distinguish between the two regimes, it is in principle necessary to perform measurements with a photon energy bandwidth smaller
than the core linewidth.

Exploiting the polarization properties of the light, the angle resolved and spin polarized detection of the decay products can allow in principle to perform highly differential experiments.
Several works have been devoted to the study of dichroism in the resonant Auger decay (with focus on the spectator channels), in
normal Auger emission and in RIXS 
\cite{mcdnormal1,mcdnormal2,mcdnormal3,cot,prloffi}. 
The different excitation
conditions in the resonant and normal Auger process result in a
different degree of polarization for the intermediate core hole
(nearly no polarization in normal Auger), consequently leading to a
different MCD. The experimental geometry that is often considered is
the so called $transverse$ or $perpendicular$ $geometry$, in which the photon
beam is perpendicular to the magnetization. In this case, the MCD in
absorption vanishes and it has been shown 
that $2p3p3p$ RPES directly displays the quadrupole moment of the
core hole \cite{quadr} (the 3d shell is merely a spectator in such decay). Decay processes involving open shells, such as 
core-core-valence and core-valence-valence decays, are more complicated 
and have not been discussed in such previous works, 
neither the photoelectron diffraction effects. Furthermore, similarly to direct photoemission/Auger
 emission from magnetic surfaces, adding spin resolution to the magnetic dichroism analysis can allow to separate different contributions to the spin polarization (SP) of the outgoing electrons \cite{sirotti,spsoandeieffects,spsoandeieffects2,hille}, but understanding the interplay between dichroism and spin polarization in autoionization channels is not straightforward.

While several theoretical formulations have recently been proposed for RIXS \cite{haverkort,kas}, similar theoretical effort is lacking for RPES.  At present, the interpretation of RPES in solids is essentially based on localized models \cite{cot,degroot} supported by multiplet calculations, with
focus mainly on the spectator channel with well defined two holes
final states. Recently, we have presented a first principle approach based on real space multiple scattering (RSMS) \cite{dapieve13}, with explicit calculations for Cr, a weak antiferromagnet. Such a method represents a practical computational scheme which allows to consider the band structure of the system, probed by the multiple scattering events felt by the excited electrons.

In this work, we reformulate our approach within a Green's function formalism and we perform calculations for spin- and angle-resolved direct valence band PES (spin resolved ARPES) and RPES (spin resolved AR-RPES) for Fe(010) for excitations at
the $L_{3}$ edge by circularly polarized light. The paper is organized as follows: in section I we present the theoretical description of the resonant process via a Green's function formalism and we give details about the calculations; in section IIA we discuss single atom results, allowing for a clear explanation of the resonance mechanism, the discrimination of intra-atomic effects in spin-flip transitions and spin polarization effects; in section IIB we move to the full cluster results, investigating the enhancement of the peaks in relation to electronic states of different spin and orbital symmetry, the spin flip transitions, and the effect of multiple scattering effects in different geometrical set up. Improvements to our approach are also discussed.

\section{Theoretical formulation}

Previous formulations of RPES are due to Davis and Feldkamp \cite{davis}, in terms
of the interaction between discrete and continuum states and by \r{A}berg et
al \cite{aberg,aberg8} in the frame of a unified theory of inelastic  scattering with the
time independent scattering theory including asymptotically the double
emission region. RPES was also reviewed by F. Gel'mukhanov and H. \r{A}gren \cite{gel}. Previous RPES calculations have been carried out with
semi-empirical methods, using either a band picture \cite{cho87,janowitz92} or
a charge-transfer cluster model \cite{tanaka,cot,degroot}.

Here we will
base our formulation on real space multiple scattering theory, excluding the region of double emission, which was previously
treated in a separate work \cite{dapieve}. We will limit ourselves to
the participator channel. A strict distinction between
such channel and the spectator one is only meaningful in the single
particle approach, which is justifiable here as we are mainly interested in
effects related to the nature of the dichroism itself or
effects related to the specific direction of the photoelectron.

The theoretical description and computational method for RPES within RSMS have been
described in our previous work on Cr(110)~\cite{dapieve13}.
For the convenience of the reader we shall outline the method
here again via a more general Green's function formulation.
In the following, $|)$ and $|\rangle$ denote
many-electron and one-electron states, respectively.
Most generally, the photoemission intensity is given by
\begin{equation}\label{eqgeneral}
I = \sum_f |(f| T|0)|^2 \delta(\hbar\omega+E_0-E_F) 
\end{equation}
where $\hbar\omega$ is the photon energy, 
$|0)$ the electronic ground state with energy $E_0$ and
$|f)$ a final state with an electron in the continuum state
$|k\rangle$ and a hole in a valence state $|v\rangle$. 
$T$ denotes the transition operator.
In the independent particle approximation (IPA), the final states are
of the form
$|f) = a^+_k a_v|0)$ where $a,a^+$ denote annihilation
and creation operators.
Using the IPA and a one-step model, 
Pendry~\cite{pendry76} showed that the (non-resonant) 
photocurrent can be written as
\begin{equation}\label{eqpendry}
I= -\frac{1}{\pi}{\rm Im}\langle
\phi|G^+(\epsilon_k)T G^+(\epsilon_v)T^+G^-(\epsilon_k)|\phi\rangle \;.
\end{equation}
Here $\phi$ is a plane wave with energy $\epsilon_k$
as observed at the electron detector. This wave is propagated into the 
crystal by the advanced single-particle Green's function 
$G^-(\epsilon_k)$, such that the total photoelectron final state 
$|k\rangle\equiv G^-(\epsilon_k)|\phi\rangle$
is a time-reversed LEED state. 
$G^+(\epsilon_v)$ is the retarded Green's function
which describes propagation of the electron inside the crystal 
with initial state energy $\epsilon_v=\epsilon_k-\hbar\omega$.
In non-resonant conditions $T$ is given by the optical
(dipole) operator $D$.
Electron correlation and life time effects may be accounted for
in a quasi-particle picture
by introducing a complex self-energy in the calculation of 
$G^+(\epsilon_v)$ and $G^-(\epsilon_k)$~\cite{braun}.

For photon energies $\omega$ around a X-ray absorption threshold,
a second transition channel opens up which leads
to the same PE final state $|f) = a^+_k a_v|0)$ 
as normal photoemission.
This resonant channel consists of a virtual X-ray absorption process
followed by an autoionization (or ``participator Auger'') decay.
To lowest order in the autoionization operator ${V}$
the transition operator $T$ then becomes~\cite{tanaka,marpe,marpe2}
\begin{equation}\label{eqT}
T(\omega) = D + \sum_m 
\frac{V|m)(m|D}{\omega + E_0 - E_m - i\Gamma_m}
\end{equation}
The sum runs over all intermediate states $|m)$
with energy $E_m$ and lifetime width $\Gamma_m$. 
Here the relevant states $|m)$ are core-excited absorption final states.
In the IPA they are of the form $a^+_ua_c|0)$, where 
$|c\rangle$ denotes a core- and $|u\rangle$ an unoccupied valence state.
Such intermediate states correspond to the initial state rule of 
X-ray absorption.
Relaxation to the core-hole may be accounted for
by calculating the orbitals $|u\rangle$ not with ground state
but with a screened core-hole potential, 
which would correspond to the final state rule of X-ray absorption.
Putting together Eqs(\ref{eqgeneral},\ref{eqT}) 
with the IPA states for $|0)$, $|m)$ and $|f)$ we obtain 
\begin{equation}\label{eqMgM}
I\sim -\frac{1}{\pi}{\rm Im}
\int dx dx' M_k(x) g^+(x,x';\epsilon_v) M_k^*(x')
\end{equation}
where $g^+(x,x',\epsilon)$ is the position representation of the
retarded Green's function $G^+$. Here $x$ is short-hand for 
$({\bf r},\sigma)$. The matrix elements are given by
\begin{equation}\label{eqMx}
M_k(x) = \langle k|D|x\rangle + 
\sum_{uc}^{\epsilon_u > \epsilon_F} 
\frac{(\langle k,c |-\langle c,k|) V|x, u\rangle \langle u|D|c\rangle}{\hbar\omega  +\epsilon_c - \epsilon_u - i\Gamma_{c}} \;.
\end{equation}
We introduce the particle Green's function $g^p(z)$ defined 
as the single-particle Green's function for
complex energy $z$ and projected on the space of unoccupied states:
\[
g^p(x,x';z) \equiv \sum_u^{\epsilon_u > \epsilon_F}
\frac{\phi_u(x)\phi_u^*(x')}{z-\epsilon_u}
\]
With $g^p$, the sum over $u$ can formally be omitted and we get 
\begin{eqnarray}\label{eqcs2}
&& M_k(x) = \phi_k^*(x)D(x) \nonumber\\
&& + \sum_{c} \int dx'dx''
\frac{\phi^*_k(x)\phi^*_c(x')-\phi^*_k(x)\phi^*_c(x')}{|x-x'|} \nonumber\\
&& \times\;
g^p(x',x'';\hbar\omega+\epsilon_c -i\Gamma_{c})D(x'')\phi_c(x'')
\end{eqnarray}
Complex-valued single-particle Green's functions 
can be computed efficiently using multiple scattering 
theory~\cite{sebilleau06}.
This theoretical approach has been implemented in a
real space full multiple scattering method. Explicit formulas of the
resonant cross-section can be found in our previous paper~\cite{dapieve13}.  Note
however, that the function $g_p$ was not used but the energy integration
over $\epsilon_u$ in Eq.~(\ref{eqMx}) was carried out numerically. The real space multiple
scattering code \cite{krueger11,dapieve13} is interfaced with self-consistent all electron potentials obtained with the band structure method LMTO. 
In this work on Fe(010), the atomic potentials were computed
in the local spin density approximation for bulk ferromagnetic Fe.
The calculated magnetic moment of 2.26 $\mu_{\rm B}$ is in good agreement 
with experiment.
The dipole and Auger matrix elements are calculated
using scalar relativistic wave functions.
From the latter, the spin-orbit coupled $2p_{3/2}$ core states are build up.
The much weaker spin-orbit coupling of the valence and continuum states
has been neglected. For the optical transitions, the dipole approximation in the              
acceleration form is used~\cite{pendry76}, since the length form is 
not well defined for delocalized state. The theoretical spectra presented in the next
sessions include a lorentzian broadening FWHM of 0.2 eV.

\section{RESULTS AND DISCUSSION}

\subsection{\label{sec:level2} Spin polarized MCD in angle resolved RPES from a single Fe atom}

First, we illustrate the resonance mechanism for the case
of emission from one single Fe atom, and we discuss the energy dependence of the signal, the origin of spin flip transitions and spin polarization.  
The interstitial potential and the potential for the absorbing Fe site are the
same as those used in the cluster calculations presented in the following paragraph. The difference 
is that all multiple scattering and thus all band structure effects
are absent.
This allows to focus on the intra-atomic origin of polarization 
and spin dependence of the resonant photocurrent. In Fig 1 we show the spin and angle resolved
direct valence band PES and RPES (spin resolved ARPES and AR-RPES) cross section for left and right circular
polarization, for four photon energies across
the $L_3$ edge and the corresponding costant initial state (CIS)
spectra ($h\nu$=680.57, 681.50, 683.83, 693.60 eV). 
The direction of the incoming beam is chosen to be
collinear with the spin magnetic 
moment (parallel geometry) while the electron is emitted in a 
perpendicular direction.

\begin{figure*}
\begin{center}
\includegraphics[clip=,width=17cm,height=8cm]{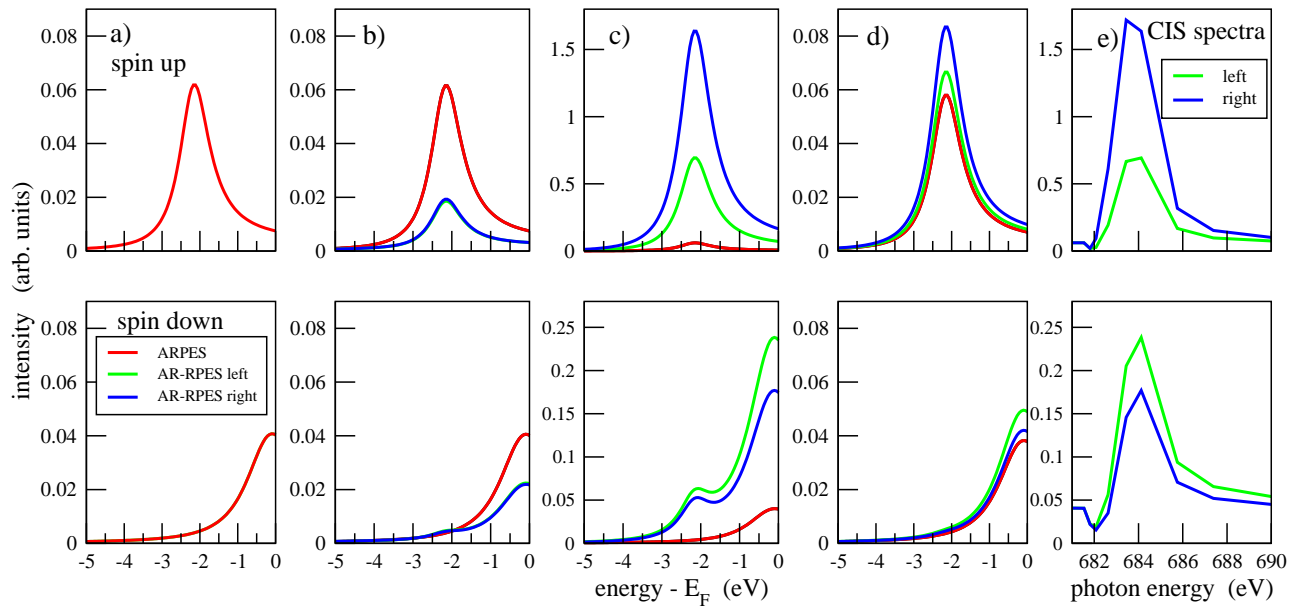} 
\end{center}
\caption{Spin and angle resolved direct valence band PES (ARPES) and RPES (AR-RPES) at the $L_3$ edge for one single Fe absorber in the single atom limit, for light incidence along the magnetic moment 
(parallel geometry) and emission in a perpendicular direction.
Left and right circular polarized light in green and blue.
Direct PES process alone (``ARPES", red) for comparison.
Photon energies are 680.57 (a), 681.50 (b), 683.83 (c), 693.60 eV (d).
(e): constant initial state (CIS) spectra at peak maxima
($E=-2.2$/$-0.2$~eV for spin up/down).}
\label{fig:oneatom}
\end{figure*}

{\it Energy dependence.} For the first photon energy the decay channel is not yet open, thus
only the direct valence band PES can take place (Fig.~\ref{fig:oneatom}(a)). The second photon energy is also
in principle below threshold, but because of finite core-hole lifetime $\Gamma_c$,
the onset of the opening of the resonant channel can occur at photon
energies slightly below such nominal threshold (Fig.~\ref{fig:oneatom}(b)). For this photon energy we observe a destructive interference between the direct and the
resonant channel, i.e. the opening of the autoionization path
decreases the total emission intensity. This corresponds to the dip region in
the CIS spectrum. For the third photon energy, the intensity is strongly enhanced by the opening of the
core hole assisted channel (Fig.~\ref{fig:oneatom}(c)), while far from the resonance the total intensity
goes back to the one corresponding to the simple direct valence band PES
process (Fig.~\ref{fig:oneatom}(d)). The CIS spectra (Fig.~\ref{fig:oneatom}(e)) show a Fano profile typical of interference
processes.

{\it Spin-flip transitions.} The direct valence band signal does not show circular dichroism.  This is
expected since for magnetic circular dichroism, spin orbit (SO) coupling is
necessary, but here such coupling is neglected in valence and continuum states.
Circular dichroism in the angular distribution
(CDAD)~\cite{jmmm,west,fecher,schon,schon2,ritchie,parzy,che} effects are also absent, as the
geometrical set up is not chiral. The RPES signal, however,
shows a large dichroism, which is maximum for the
strongest enhancement of the signal and of opposite sign for the two
spin channels.  Since any band structure effects are absent in this
single atom limit, the direct signal displays a single peak
of lorentzian shape typical for a $d$-wave potential scattering resonance
and spin up and spin down peaks separated by the exchange splitting.
At resonance however, a new feature shows up in the spin down channel
corresponding to the energy of the peak of the spin up channel (-2.2 eV). 
As there are (almost) no spin down valence states at this
energy, this peak in the spin down photocurrent corresponds to spin up
initial states. This means that the spin of the photoelectron is
opposite to the one of the final valence hole, and thus it is a spin
flip transition. 

In a spin flip process, the selection rule $\Delta
S=0$ is lifted, since, due to considerable core SO coupling, spin is not
a conserved quantity. The origin of such spin-flip transitions, both in resonant Auger and 
RIXS \cite{laub,kachel,gallet,dallera,chiuzbaian,braicovich,gerken},
has always been at the centre of a debate about whether they take place 
in the absorption step or in the core-hole decay.
The spin flip transitions observed here
are a combined effect of spin-orbit coupling in the $2p_{3/2}$
core shell and exchange Coulomb decay.
It is clear that direct Coulomb decay cannot give rise to spin-flip
transitions, since the matrix element is
$\langle k\sigma c\sigma'|V|v\sigma u\sigma'\rangle$
and so the photoelectron $k$ and the valence hole $v$ have necessarily the 
same spin.
However, in exchange decay with matrix element
$\langle c\sigma k\sigma'|V|v\sigma u\sigma'\rangle$
 spin-flip can occur for $\sigma=-\sigma'$.
As the corresponding dipole transition matrix element is 
$\langle u\sigma'|D| c\sigma'\rangle$, the process also
requires a spin flip of the core hole
$|c\sigma'\rangle\rightarrow |c\sigma\rangle$.
This is only possible when the core eigenstates 
have mixed spin character due to SO coupling, 
which is the case for the $2p_{3/2}$ $m_j=\pm 1/2$
states.

{\it Spin polarization.} The peak intensity ratio between up and down spin is about 5:1 at 
maximum resonance 
(Fig.~\ref{fig:oneatom}(c), averaging over the two light polarizations)
while it is only about 3:2 off resonance (Fig.~\ref{fig:oneatom}(a)).
So the resonant process leads to a large enhancement of the 
valence band spin polarization. 
%
%
This can be understood as follows.
Since the spin up band is almost full, core-valence excitation
can only happen for spin down electrons. This means that
the large majority of intermediate states are spin down particle-hole 
excitations $|m)= a^+_{u\downarrow}a_{c\downarrow}|0)$.
The autoionization decay of such intermediate states gives rise
to both spin up and spin down electrons, but with very different
transition probabilities.
Spin up photoelectrons $|k\uparrow\rangle$ can only be produced
through direct Coulomb decay, whose matrix element is 
$V_D(\uparrow)=\langle k\uparrow c\downarrow|V|v\uparrow u\downarrow\rangle$.
For spin down electrons the direct matrix element is
$V_D(\downarrow)                                                               
=\langle k\downarrow c\downarrow|V|v\downarrow u\downarrow\rangle$
and the corresponding transition probability is 
smaller by a factor $n(v\downarrow)/n(v\uparrow)$,
where $n$ are the ground state occupation numbers.
This is the same ratio as in the direct (non-resonant) photoemission process.
So if there were only direct Coulomb decay, resonant and non-resonant
photoemission would have the same degree of spin polarization.
The observed resonant enhancement of the spin polarization is due
to the exchange decay. From intermediate states of the form
$a^+_{u\downarrow}a_{c\downarrow}|0)$, exchange decay produces only 
spin-down electrons, with matrix element
$V_X(\downarrow)=\langle c\downarrow k\downarrow|V|v\downarrow u\downarrow\rangle$. The total decay matrix element for spin down photoelectrons
is $V(\downarrow)=V_D(\downarrow)-V_X(\downarrow)$ (see Eq.~\ref{eqMx}).
Now $V_X(\downarrow)$ is comparable with $V_D(\downarrow)$,
since the radial matrix elements are exactly the same when 
$|u\rangle$ and $|v\rangle$ are both $3d$ states.
Thus the exchange decay strongly reduces the spin down 
transition amplitude with respect to the direct Coulomb decay alone. 
For spin up photoelectrons, however,
no such reduction occurs, because $V_X(\uparrow) = 0$.
This explains why the resonant process produces much more spin up
than spin-down electrons.

Note that already in our recent study on RPES from Cr~\cite{dapieve13}
we have argued that a core-valence excitation of one spin channel leads,
through autoionization decay, to a majority of photoelectrons with the
opposite spin. It is a crucial element in the explanation why the 
so-called mixed signal is non-zero even for a non-magnetic ground state. This issue will be discussed here below, now for the case of a strong ferromagnet.

{\it Interplay between dichroism and spin-polarization.}
When using left (+) or right ($-$) circular polarized light and 
spin-resolution ($\uparrow$,$\downarrow$) of the photoelectrons,
there are four independent spectra.
We consider the following ``fundamental'' combinations:\\

\begin{tabular}{lccl}
tot & = & $(+\uparrow) + (-\uparrow) + (+\downarrow) + (-\downarrow)$ &(total)\\
dic & = & $(+\uparrow) - (-\uparrow) + (+\downarrow) - (-\downarrow)$ &(dichroic)\\
spr & = & $(+\uparrow) + (-\uparrow) - (+\downarrow) - (-\downarrow)$ & (spin-polarized)\\
mix & = & $(+\uparrow) - (-\uparrow) - (+\downarrow) + (-\downarrow)$ & (mixed)\\
\end{tabular}

\begin{figure}[!htb]
\begin{center}
\includegraphics[clip=,width=0.82\columnwidth]{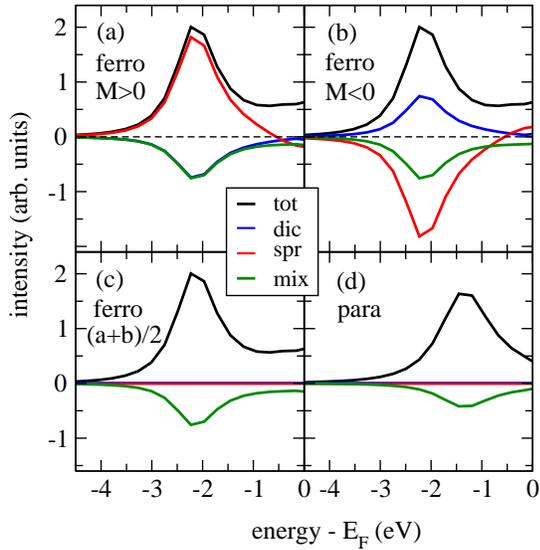}
\end{center}
\caption{(Color online). 
RPES at maximum resonance ($h\nu$=683.83 eV)
for a single Fe atom. Same geometry as in Fig.~1.
Fundamental spectra total (tot), dichroic (dic), spin-polarized (spr) 
and mixed (mix) for ferromagnetic (a-c) or Pauli para-magnetic 
(d) ground state.}
\label{fig:check} 
\end{figure}

In Fig.~\ref{fig:check}(a) we plot these fundamental spectra for a single Fe atom at
maximum resonance. The set-up is the same as in Fig. 1 with light incidence 
along the magnetization axis (+z) and electron emission perpendicular 
to it (+y). 

The dichroic signal in Fig.~\ref{fig:check}(a) is large and negative, which is a direct
consequence of the negative circular dichroism in X-ray 
absorption at the $L_3$ edge, which enters here as the excitation step
in the resonant process. 
As expected from the direct valence band spectra in Fig.1a,
the spin polarization changes sign between the majority spin peak at 
-2.2~eV and the minority spin peak around $E_F$.
In resonant conditions, the majority spin peak is, however, 
much more enhanced than the minority peak (as discussed before)
such that the spin-polarized spectra is
dominated by the positive majority peak. 
The mixed signal is large and negative. It closely follows 
the dichroic signal along the majority peak, but stays negative at 
the minority peak contrary to the dichroic signal which becomes
negligible around $E_F$.
In Fig.~\ref{fig:check}(b) the magnetization direction is reversed ($M<0$).
As expected from their symmetry under time reversal, both dichroic and 
spin-polarized spectra change sign, while the mixed signal remains 
unchanged~\cite{vdlcomment}.
This confirms that the mixed signal analyzed in some earlier pioneering studies ~\cite{sinkovic,tjeng} is essentially independent of the orientation
of the magnetic moments. 
In Fig.~\ref{fig:check}(c) we have plotted the 
average of the spectra in (a) and (b), meant as a simple model 
for a ferromagnet with vanishing macroscopic magnetization due
to disordered moments or domain structure.
Clearly, the spin polarized and dichroic signals vanish, 
but the mixed signal does not, as found experimentally for 
Ni above the Curie temperature~\cite{sinkovic}.
In Fig.~\ref{fig:check}(d) we show the fundamental spectra obtained for Fe
with a non-magnetic ground state, which would correspond to a Pauli 
paramagnetic system.
As exchange-splitting is absent, the spectrum consists of a single broad
peak centered around -1.3 eV. For this non-magnetic  
system and non-chiral set up, the dichroic and spin-polarized signals are obviously zero.
However, the mixed signal is of the same sign and order of magnitude as 
that found in the ferromagnetic system (a-c). 
This shows that the mixed signal is mainly of non-magnetic origin.

We have drawn the same conclusion previously in the case of 
Cr~\cite{dapieve13}, i.e. for a weak antiferromagnet.
In that case, the mixed signal was found almost identical 
for the antiferromagnetic to the paramagnetic 
ground state~\cite{dapieve13}.
In the present case of the Fe atom with large magnetic moment and 
exchange splitting, the mixed signal clearly 
changes both in position and amplitude when going from the magnetic
(Fig.~\ref{fig:check}a-c) to non-magnetic ground state (Fig.~\ref{fig:check}d).
Qualitatively the same changes are, however, observed for the total
spectrum (tot), which means that the mixed signal does not yield
more information about the magnetic state of the system than the
total (isotropic) spectrum.
Thus our main conclusion from the Cr results is confirmed here
for a ferromagnetic system with large moments: the mixed signal 
is not due to the presence of local magnetic moments, but rather
reflects the spin-orbit coupling of the $2p_{3/2}$ shell, which is
``transferred'' to RPES through to 
the exchange process of the autoionization decay~\cite{dapieve13}.
Interestingly, a sensitivity to local magnetic properties above and below the transition temperature has been reported in itinerant ferromagnets by spin unpolarized angle-resolved coincidence detection of the photoelectron and the Auger electron in the normal Auger decay \cite{prloffi}. 

 \subsection{\label{sec:level2} Fe(010) RPES in parallel geometry and normal emission }

We now move to the analysis of spin and angle resolved
RPES from a Fe cluster. We start from the case of parallel geometry with normal emission for the outgoing electrons, and we discuss the energy dependence of the signal, the sensitivity to electronic states of different spin and orbital sysmmetry and spin flip transitions.

\begin{figure}[!htb]
\begin{center}
\includegraphics[clip=,width=3.8cm,height=4.0cm]{confrontoDOSnew.eps}
\includegraphics[clip=,width=4.2cm,height=3.9cm]{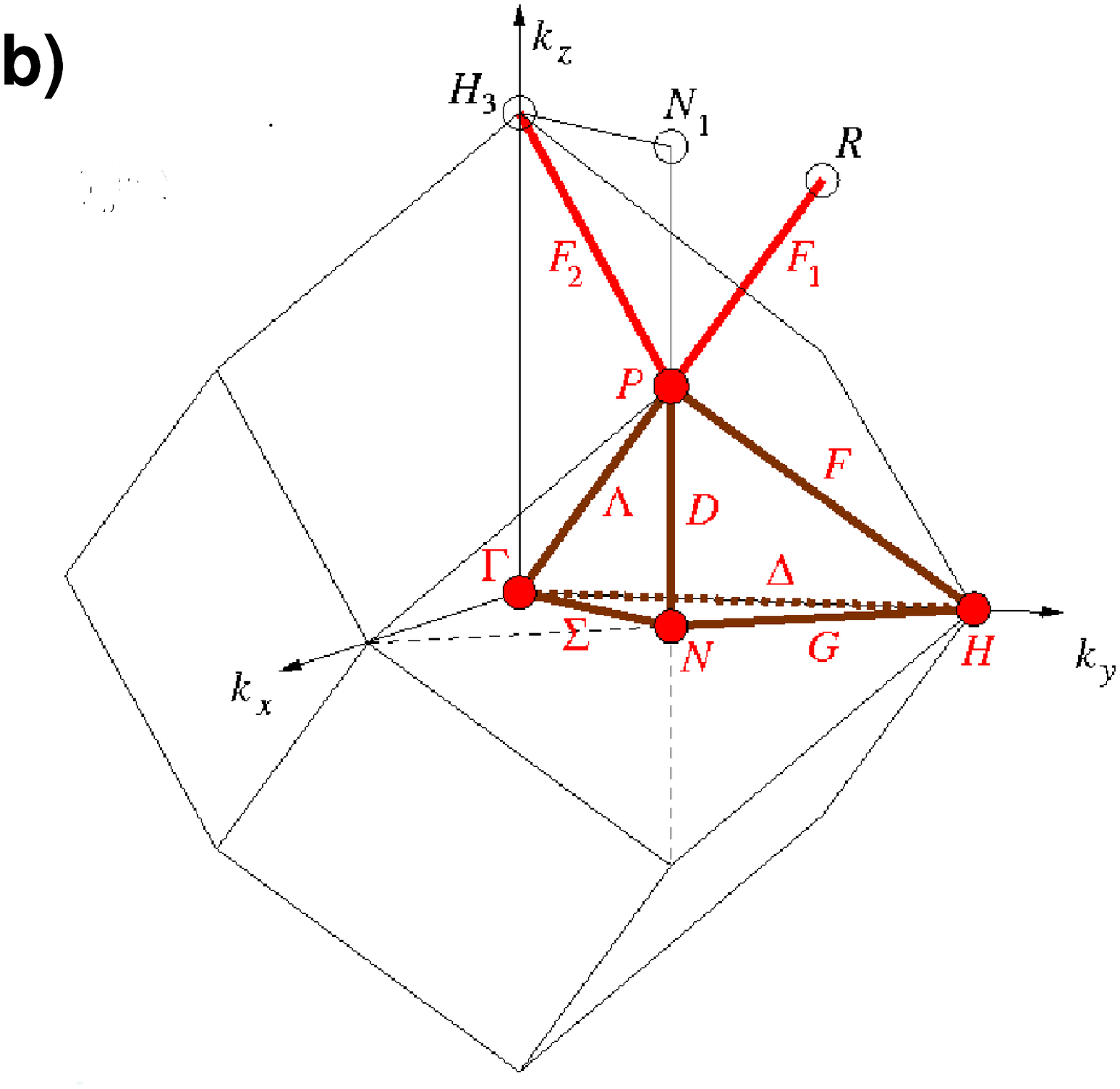}\\
\includegraphics[clip=,width=0.485\columnwidth,height=4.5cm]{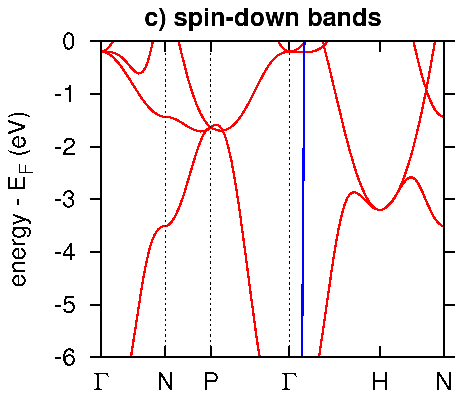}
\includegraphics[clip=,width=0.485\columnwidth,height=4.5cm]{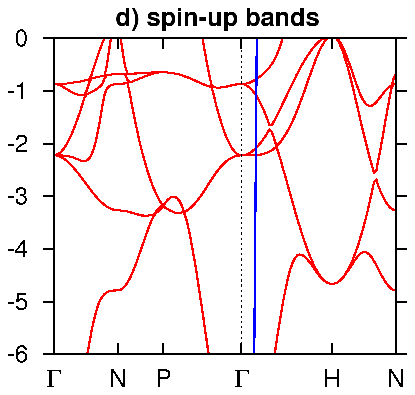}
\end{center}
\caption{a) DOS of bulk bcc Fe from
periodic LMTO calculation compared with DOS on central atom 
in the cluster in RSMS calculation (positive/negative DOS for spin-up/down); b) bcc Brillouin zone; c) spin-down
bands from LMTO calculation (red) and final state free electron dispersion
(blue); d) the same as c) now for spin-up bands.}
\label{fig:bs}
\end{figure}

The Fe(010) surface is modeled with a semi-spherical
 cluster of 184 atoms and the magnetization is 
 assumed in-plane along $<$001$>$. In Fig.~\ref{fig:bs}(a) we show the comparison between the density of states (DOS) calculated by LMTO on bulk ferromagnetic Fe and our RSMS
  code for a central atom in the cluster. The agreement between the DOS by LMTO and the DOS for a central atom in
  the cluster is very good, showing that the bulk electronic properties are
  well described by the internal atoms of the cluster. Fig.~\ref{fig:bs}(c,d) show the band structure along the high symmetry lines
of the bcc Brillouin zone of bulk Fe Fig.~\ref{fig:bs}(b), as calculated with the LMTO code.

 The reference
frame attached to the cluster is such that the $z$ axis is defined by
the magnetization direction, with the magnetic moment pointing along
$+z$, the $y$ axis is perpendicular to the surface and the $x$ axis is still
lying on the surface. In the case of parallel geometry, the light incidence
direction is along $+z$.

{\it Energy dependence and MCD}. In this geometry, the MCD in XAS is maximum, since 
it essentially measures the projection of the magnetic 
moment onto the direction of the light incidence.
In Fig.4 we show the spin resolved ARPES and AR-RPES
intensities for the Fe(010) cluster for left and right circular
polarization, again for four photon energies across the $L_3$ edge. In this geometrical set up, there is no source for additional purely geometric dichroism
(CDAD), since all the relevant vectors are coplanar (and along high
symmetry directions) and hence there is no
chirality induced solely by the experimental set up. We again observe a region of
deconstructive interference (Fig.4(b)) and then a strong enhancement
for the third photon energy (Fig.4(c)), which is different for the two
spin channels. The massive enhancement of the signal observed here
does not imply strong interference effects: an analysis of the different contributions in the amplitude reveals that, in our case, the
enhancement is given essentially  by the resonant
excitation alone, as was also found in other cases \cite{magnuson}.

\begin{figure*}[!htb]
\parbox{0.38\columnwidth}{
\begin{center}
\includegraphics[clip=,width=0.38\columnwidth]{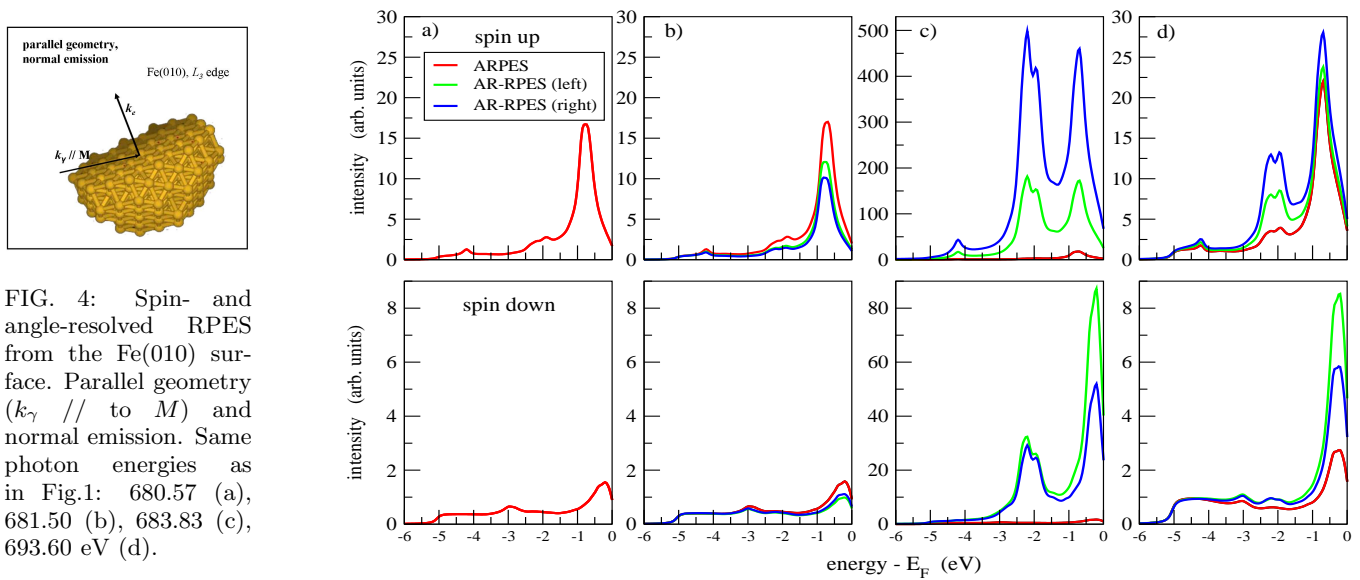}\\
\caption{Spin- and angle-resolved RPES from the Fe(010) surface. Parallel geometry ($k_\gamma$ // to $M$) and normal emission. Same photon energies as in Fig.1: 680.57 (a), 681.50 (b), 683.83 (c), 693.60 eV (d).}
\end{center}
}
\parbox{1.55\columnwidth}{
\begin{center}
\hspace*{3.5em}\includegraphics[clip=,width=1.55\columnwidth,height=7.6cm]{normemiss_parallelgeom_novinner_sm_bis.eps}
\end{center}
\label{fig:geom1}
}
\end{figure*}

As compared to the single atom spectra in Fig.1, the
cluster spectra show various new features due
to electron scattering (discussed below in connection 
with the band structure of the system).
However, in the geometry considered here, electron scattering
does not seem to act as an additional source of dichroism,
since the sign and shape of the dichroic signal is essentially
the same as in the single-atom case.

{\it Sensitivity to orbital symmetry and spin flip transitions}. Let's now discuss the sensitivity of the ARPES and AR-RPES spectra to electronic states of different orbital symmetry. We shall first discuss the spin up channel.
The flat band along $\Gamma$-N-P-$\Gamma$ around -1~eV in Fig.~\ref{fig:bs}(d) 
gives rise to the strongest peak in the DOS as well as in the non 
resonant photoelectron spectrum at normal emission.
A projection of band states onto atomic orbitals (not shown) reveals that 
the flat band at -1~eV is essentially of $e_g$ character. 
The two states at the $\Gamma$ point, at                
-0.9 and -2.2 eV, are of pure $e_g$ and $t_{2g}$ character, respectively.
From $k_{||}$ conservation it follows that for normal emission
($k_{||}$=0) the initial states lie on the $\Gamma$-H line.
The peak positions of the spectrum can be found by plotting 
the final state bands downshifted by the photon energy. The
crossing points give the possible direct optical transitions in bulk Fe.
Assuming free electron dispersion, we have plotted the shifted final state
band for a photon energy of 683.8~eV and normal emission ($k_{||}=0$)
as a blue line in Fig.~\ref{fig:bs}(c,d).
The crossing points are close to the $\Gamma$ point, because
for $E$(initial)=0, we have $k$(final)$=\frac{2\pi}{a} (0,0,6.16)$
and $\Gamma$ (H) points are at even (odd) 
multiples of $\frac{2\pi}{a} (0,0,1)$.
Note that since the slope of the final state parabola is very large,
a moderate change of the photon energy leads to only a small horizontal
shift of the blue line, e.g. by 4\% of the $\Gamma$-H distance for 
a photon energy change of 10~eV.
It can be seen that the crossing points fit quite well the photoemission 
peaks around -0.8 and -2.0 eV confirming the band mapping interpretation
of the valence band PES.
Weak extra peaks (e.g. at -4.2~eV) may be due
to umklapp processes which can lead to different crossing points on 
the $\Gamma$-H line.
 
It is interesting to note that in off resonance conditions (Fig.4(a)) 
the peak at -2.0~eV which corresponds to an initial state of 
$t_{2g}$ character, is much weaker than the $e_g$ peak at -0.8~eV.
This can essentially be understood from orbital selection rules.
In a reference frame where the surface normal is chosen as the
z-axis, only $m_l=0$ final states contribute to normal emission.
From angular momentum recoupling coefficients and 
dipole selection rules it is then straightforward 
to show that for the dominating $d$ to $f$ transitions and the
chosen light incidence,
an initial $e_g$ orbital leads to a three times larger normal 
emission intensity than a $t_{2g}$ orbital.
This argument, holds, however, only for the direct process, where the
valence state symmetry together with the optical dipole selection rule 
essentially determines the angular distribution of the photoelectrons.
For the autoionization process however, the selection rules are
more complex and involve also the symmetry of the core hole and 
excited state~$u$. This might explain why at maximum resonance,
where the autoionization process completely dominates, 
the $t_{2g}$ peak at -2.0~eV is no longer suppressed, but
of comparable strength as the $e_g$ peak at -0.8~eV (Fig.4(c)).

Turning now to the spin down channel, band mapping analysis predicts
a single normal emission peak close to $E_F$ 
(crossing point in Fig.~\ref{fig:bs}(c)). The strongest peak is indeed observed at -0.2~eV. 
When going from the non-resonant (Fig.4(a)) to the resonant spectrum
(Fig.4(c)) a new peak appears around -2.3~eV. This is clearly
a spin-flip peak, since its position and shape exactly match 
the largest peak of the resonant spin up spectrum (Fig.4(c), upper panel).
This shows that the spin-flip transitions, identified above in
the single atom case, must also be expected in AR-RPES 
from surfaces.


\begin{figure*}[!htb]
\begin{center}
\hspace*{1.8em}\includegraphics[clip=,width=0.4\columnwidth,height=3.4cm]{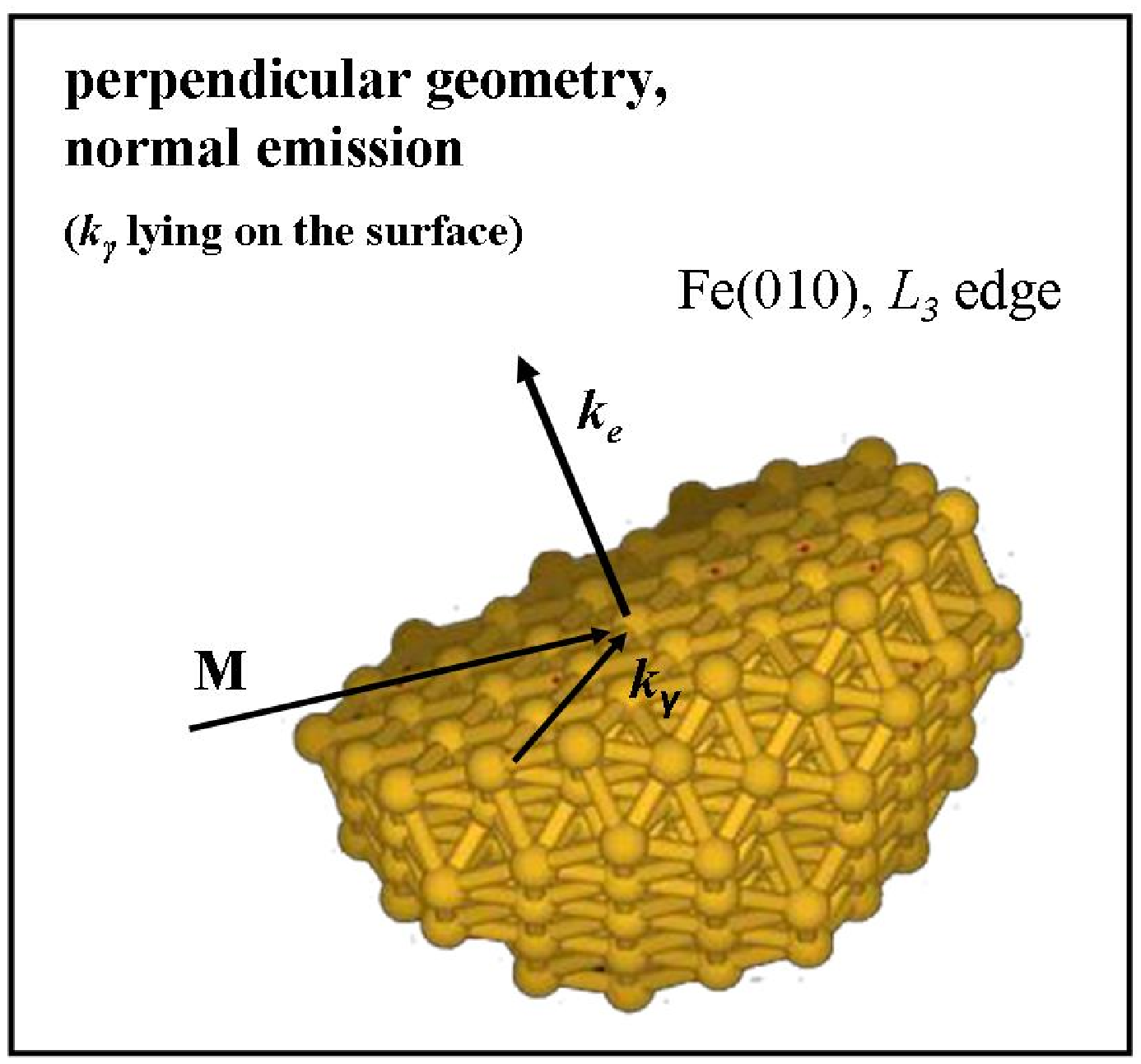}
\hspace*{3.4em}\includegraphics[clip=,width=0.4\columnwidth,height=3.4cm]{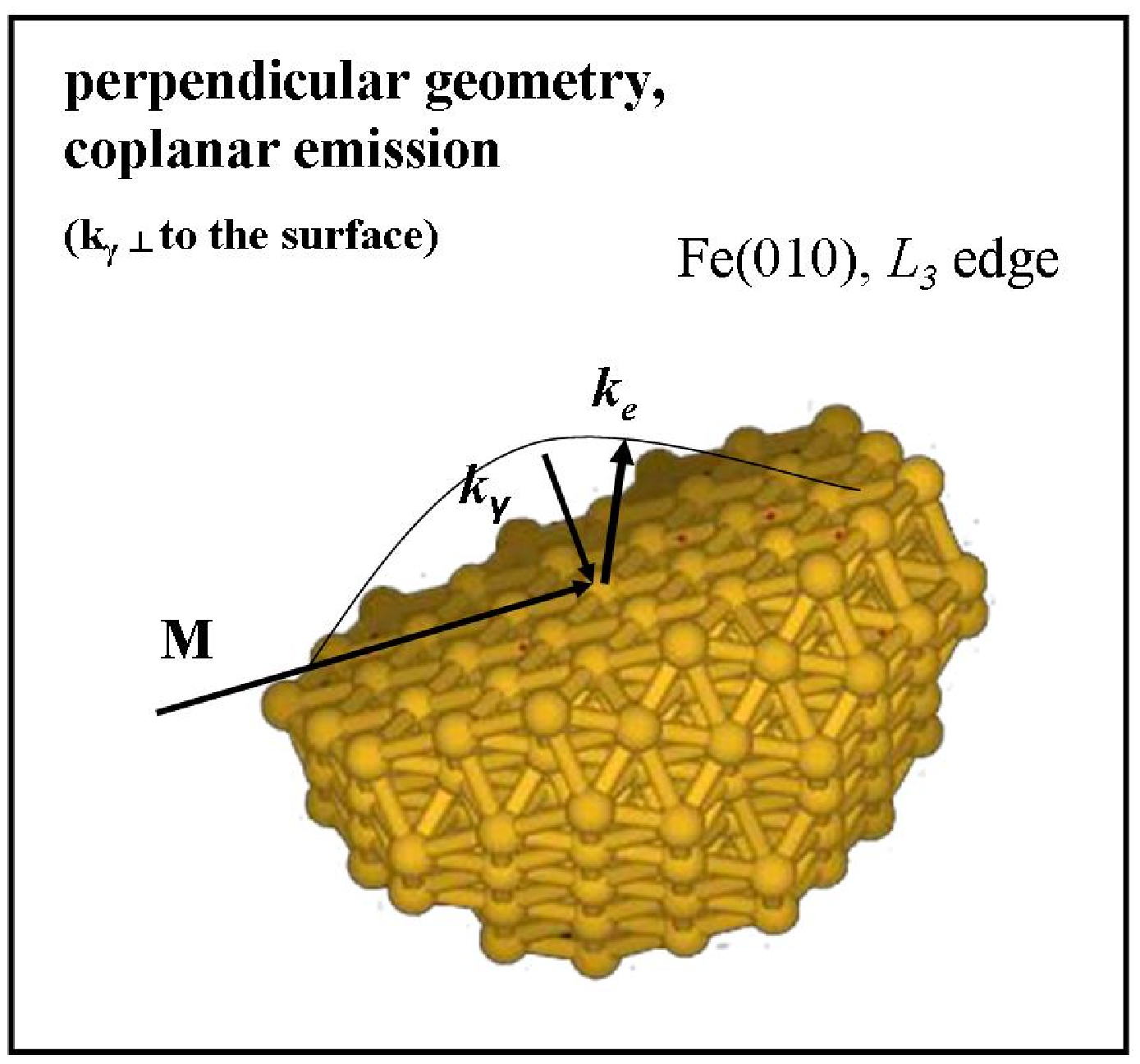}
\hspace*{3.4em}\includegraphics[clip=,width=0.4\columnwidth,height=3.4cm]{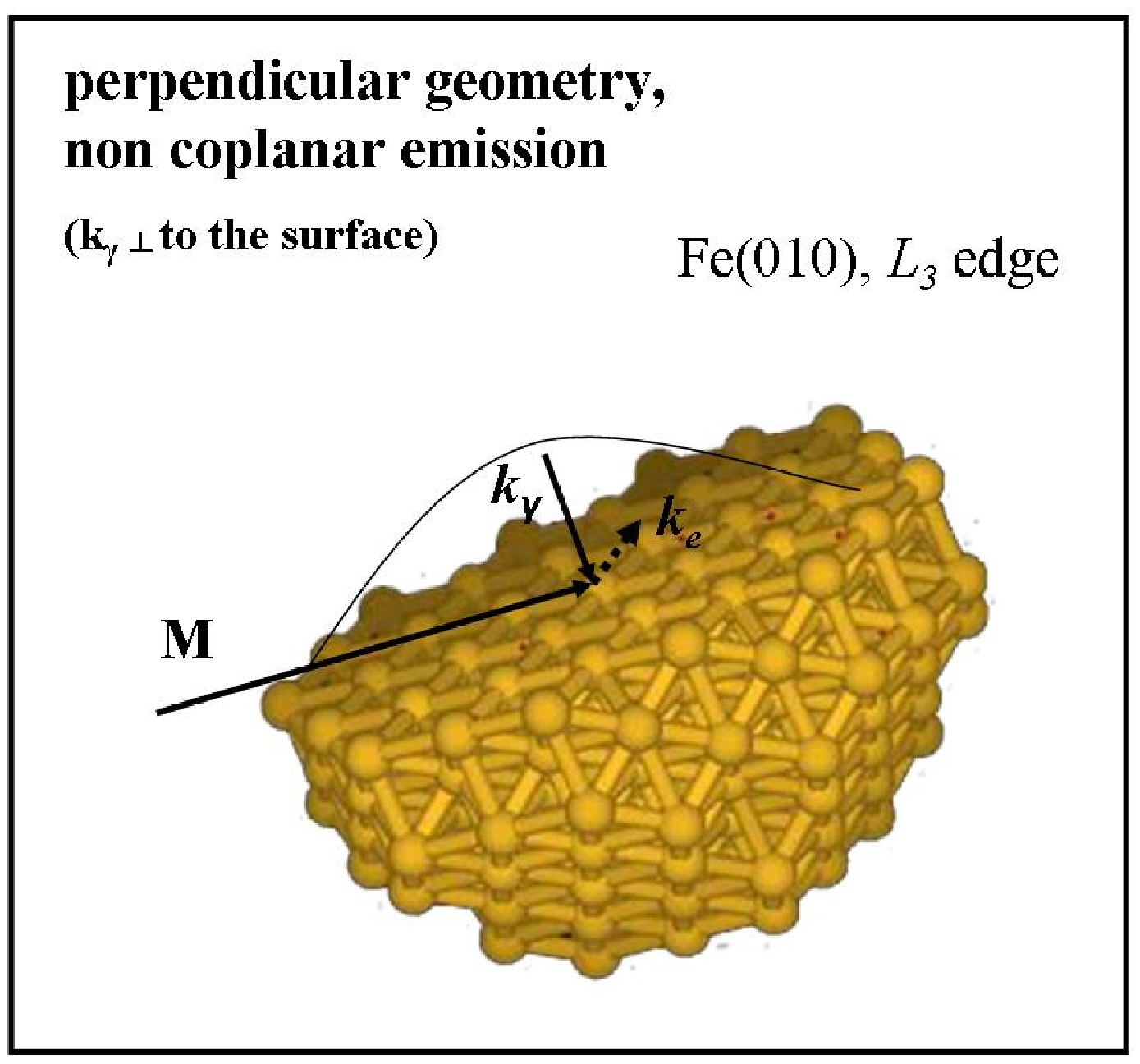}\\
\includegraphics[clip=,width=0.8\textwidth,height=8.8cm]{new_fig_geometrie_perp.eps} 
\end{center}

\caption{Spin-resolved ARPES and AR-RPES from the Fe(010) surface
in perpendicular geometry ($k_\gamma$ perpendicular to $M$) at maximum resonance
($h\nu$=683.83 eV). (a) grazing incidence (along $x$), normal emission. (b) normal incidence (along $y$), off-normal emission at ($\theta$=150$^{\circ}$,$\phi$=90$^{\circ}$) polar angles, coplanar with $M$ and $k_\gamma$. (c) normal incidence, off-normal emission at ($\theta$=150$^{\circ}$,$\phi$=65$^{\circ}$) non-coplanar with $M$ and $k_\gamma$.}
\label{fig:perpgeom}
\end{figure*}

 \subsection{\label{sec:level2} Fe(010) RPES in perpendicular geometry: 
normal and off normal emission}

We now move to the analysis of the often used $perpendicular$ or $transverse$ $geometry$, in which the signal allows to probe directly the core hole polarization \cite{quadr} avoiding the contribution of the MCD due to the absorption step (which vanishes in this set-up). Here, we consider RPES at maximum resonance and discuss circular dichroism
and spin polarization of the photoelectrons 
as a function of emission direction. We analyze two types of perpendicular geometry: the first in which the photon beam direction lies on the surface and the second in which the beam is incoming perpendicular to the surface. 


{\it Perpendicular geometry, normal emission}.  In Fig.~\ref{fig:perpgeom}(a) we
show the direct and resonant signals for the $L_3$ edge at the maximum of the resonance. $k_\gamma$ lies on the surface, and in particular it is along $+x$, thus perpendicular to the magnetization. The
magnetization is thus perpendicular to the scattering plane. The signal is spin polarized, as expected, but the amount of spin polarization, i.e. the ratio between the spin up and spin down intensity, is $\sim$3 here, and thus it is strongly different with respect to the case of parallel geometry (where it was $\sim$5). The dichroism, on the other hand, is null. This is both because in perpendicular geometry MCD in the
  first absorption step is forbidden and because CDAD effects are absent (it is a completely
  orthogonal geometry, i.e., all three relevant vector are orthogonal to
  each other and the emission is along a high symmetry direction \cite{jmmm,henk}). The symmetry
  between the two mirror conditions obtained by reversing the
polarization of the light is indeed not broken when the magnetization is perpendicular to the plane of measurement \cite{kuch,finazzi}. 
The absence of dichroism in this geometry agrees with predictions from
the atomic theory by Thole and van der Laan.
Indeed the geometrical factors $U(P,\epsilon,M)$ in
  Table I of Ref. \cite{quadr} vanish when $\epsilon$ (which denotes the photoelectron direction) is perpendicular to
$M$ and $P$ (which denotes $k_\gamma$).

{\it Perpendicular geometry, off-normal emission}. Most MCD experiments in both RPES and RIXS have been carried out
in perpendicular geometry with X-rays impinging along the surface normal. In Fig.~\ref{fig:perpgeom}(b,c), we show the spin resolved ARPES and AR-RPES
intensities for the maximum of the resonance at the $L_3$ edge in the
case of such perpendicular geometry, for two different off normal emission
directions. We have analyzed both the cases in which the electron emission direction is
coplanar with $M$ and the photon direction $k_{\gamma}$ (Fig.~\ref{fig:perpgeom}b) and
the case in which it is not (Fig.~\ref{fig:perpgeom}c).

The coplanar case is similar to
the one chosen in \cite{finazzi,vdljes}. The electron is emitted at polar angles ($\theta$,$\phi$)=(150$^{\circ}$,90$^{\circ}$)
in the Fe(010) reference frame where the surface normal
is at (90$^{\circ}$,90$^{\circ}$) and $M$ at (0$^{\circ}$,0$^{\circ}$).
Since the magnetization, photon incidence and electron emission are coplanar, there is no influence of CDAD in this set up (Fig. 5(b)) and thus only dichroism induced by the core hole polarization is probed. The amount of spin polarization is strongly reduced with respect to the geometries considered before (as in the non coplanar case discussed below), revealing that emission along non high symmetry directions strongly influences the degree of spin polarization in the photocurrent. However, still the spin up channel is subjected to a relevant enhancement with respect to the spin down channel, the origin of this has been already discussed in section IIA.
Interestingly, in the coplanar emission case, we observe that, in our approach,  the peak related to the $e_g$ is suppressed for the spin up channel, for both light helicities, suggesting that the orbital symmetry of the levels probed by the electron excited in the intermediate state can strongly influence the lineshape. Furthermore, as the core hole polarization is also determined by the population in the magnetic sublevels of the $d$ band, other ground state moments (not considered here) than the spin moment should also be taken into account, as they could play a role in determining the lineshape at the resonance, by contributing in the enhancement or suppression of specific features.

In the non coplanar case the electron is emitted at (150$^{\circ}$,65$^{\circ}$) (Fig 5(d)) and one has the combined presence of both CDAD and dichroism induced by the core hole polarization. Indeed, a non-zero (albeit weak)
circular dichroism appears even in the direct signal, because of purely geometric (CDAD) effects. Contrary to what observed
before in the case of parallel geometry and normal emission, the dichroism of the RPES signal in the two spin channels for the peak near $E_F$ (of $e_g$ and $t_{2g}$ character for the spin up and down channel respectively) is essentially of the same sign. Thus, in the case of a perpendicular geometry 
and off-normal emission directions, multiple scattering effects can even lead to a sign reversal of the spin polarized MCD signal in correspondance to certain spectral features, as reported in previous works on both core and valence direct photoemission \cite{henk,prb506774,chasse,schneider,ven2}. 
As a general trend, our results indicate that, when combining perpendicular geometry with off normal emission directions, scattering effects do considerably influence the intensities, the
dichroism, and photoelectron spin polarization.

\subsection{Limitations of the method and possible future refinements}

The present method is a fast and simple ab-initio theory
of angle-resolved RPES. As it is the first method of this kind,
it contains several assumptions and approximations. The latter might 
limit its accuracy and range of applicability and they 
should therefore be overcome in the future.
First we have limited ourselves to the participator process.
This obviously leads to problems for interpretation of
experimental data if the (Raman-like) participator spectrum 
can not easily be separated from the (Auger-like) spectator
spectrum. Second, the theory is formulated within the independent particle picture,  which implies that it should give best results for weakly correlated systems. For Fe, relevant $3d$ electron correlations and a partial breakdown of the one 
electron approach have been reported from measurements of unexpected 
magnetic dichroism in the transverse geometry in normal Auger spectroscopy
\cite{mcdnormal2} and from the description of real and imaginary parts
of the quasiparticle self energy \cite{prb85licht}. Also, it is likely that the resonance enhancement of both the total and 
spin polarized spectra are somewhat overestimated in the present approach,
because both the addition of the spectator channel and of many-body excitations
would renormalize the single-particle participator response. 
For example core-hole screening leads to an increase of the local valence 
charge and, generally, to a decrease of the valence spin polarization \cite{screening1,mcdnormal1}.
Finally, the same Fe bulk potential has been used here 
for all atoms in the cluster, for simplicity. 
This may be easily improved by taking 
layer-dependent potentials from a self-consitent surface calculation. 
Thereby the change of local magnetic moment at the surface due to valence 
bands narrowing could be taked into account \cite{sirotti,spsoandeieffects,surfbulk}. 
While this effect is rather small in
Fe(010), it might be important for other compounds and less dense surfaces.

\section{\label{sec:level2} CONCLUSIONS}

In summary, we have presented a first-principles method for 
valence band angle resolved resonant photoelectron spectroscopy (AR-RPES)
in a real space multiple scattering approach.
We have studied the spin resolved ARPES and AR-RPES at the Fe $2p_{3/2}-3d$ resonance, focussing on 
circular dichroism and spin polarization of the photoelectrons 
emitted from a Fe(010) surface in various experimental geometries.
Our results fully agree with qualititative predictions that can be gained
from general symmetry considerations, atomic models, and the band structure 
of the system. While the AR-RPES spectra reflect the band structure in 
terms of peak positions, the relative peak intensities deviate considerably
from non-resonant ARPES. Moveover spin flip transitions can lead to new
peaks in the spin resolved AR-RPES. 
The results on fundamental spectra confirm the conclusion drawn previously in the case of 
antiferromagnets \cite{dapieve13},
that the so-called mixed signal of combined circular polarized light and 
spin resolution is essentially unrelated to the existence of local 
magnetic moments. 
By comparing different geometries we have also analyzed the
influence of multiple scattering effects on dichroism and spin
polarization for emission along high and low-symmetry directions.
At present, our method offers a fast and simple ab-initio theory,
which can provide useful information about local properties of low
correlated systems, the estimation of the parameters of
electron-electron and spin-spin interactions in the resonant decay, and can provide guidelines for future experiments.

\begin{acknowledgments}

F. Da Pieve ackowledges financial support from the VUB, Free
University of Brussels, through the GOA77 project.

\end{acknowledgments}

\nocite{*}

\end{document}